\documentclass[twocolumn,noshowpacs,aps,pra,superscriptaddress ]{revtex4-1}
\usepackage{graphicx}
\usepackage{amsmath}
\usepackage{amsfonts, amsmath, amssymb, bm, graphicx}
\graphicspath{{pictures/}}
\DeclareGraphicsExtensions{.pdf,.png,.jpg,.eps}
\usepackage[breaklinks=true,unicode=true,urlcolor=blue,
colorlinks=true,citecolor=blue,linkcolor=blue]{hyperref}
\def\<{\left\langle}
\def\>{\right\rangle}

\begin{document}

\title{THz-Frequency  Spin-Hall Auto-Oscillator Based on a Canted Antiferromagnet}

\author{O.~R.~Sulymenko}
\author{O.~V.~Prokopenko}
\affiliation{Faculty of Radio Physics, Electronics and Computer Systems, Taras Shevchenko National University of Kyiv, Kyiv 01601, Ukraine}

\author{V.~S.~Tiberkevich}
\author{A.~N.~Slavin}
\affiliation{Department of Physics, Oakland University, Rochester, MI 48309, USA}

\author{B.~A.~Ivanov}
\affiliation{Faculty of Radio Physics, Electronics and Computer Systems, Taras Shevchenko National University of Kyiv, Kyiv 01601, Ukraine}
\affiliation{Institute of Magnetism, National Academy of Science of Ukraine, Kiev, 03142, Ukraine}

\author{R.~S.~Khymyn}
\affiliation{Department of Physics, University of Gothenburg, 41296 Gothenburg, Sweden}

%%%%%%%%%%%%%%%%%%%%%%%%%%%%%%%%%%%%%%%%%%%%%%%%%%%%%%%%%%%%%%%%%%%%% antiferromagnet (AFM)

\begin{abstract}
We propose a design of a THz-frequency  signal generator based on a layered structure consisting of a current-driven platinum (Pt) layer and a layer of an antiferromagnet (AFM) with easy-plane anisotropy, where the magnetization vectors of the AFM sublattices are canted inside the easy plane by the Dzyaloshinskii-Moriya interaction (DMI).  The DC electric current flowing in the Pt layer creates, due to the spin-Hall effect, a perpendicular spin current that, being injected in the AFM layer, tilts the DMI-canted AFM sublattices out of the easy plane, thus exposing them to the action of a strong internal exchange magnetic field of the AFM. The sublattice magnetizations, along with the small net magnetization vector  $\textbf{m}_{\rm DMI}$ of the canted AFM, start to rotate about the hard anisotropy axis of the AFM with the THz frequency proportional to the injected spin current and the AFM exchange field. The rotation of the small net magnetization  $\textbf{m}_{\rm DMI}$ results in the THz-frequency dipolar radiation that can be directly received by an adjacent (e.g. dielectric) resonator. We demonstrate theoretically that the radiation frequencies in the range $f=0.05-2$~THz are possible at the experimentally reachable magnitudes of the driving current density, and evaluate the power of the signal radiated into different types of  resonators, showing that this power increases with the increase of frequency $f$, and that it could exceed 1~$\mu$W at $f \sim 0.5$~THz for a typical dielectric resonator of the electric permittivity $\varepsilon \sim 10$ and quality factor $Q \sim 750$.
\end{abstract}

\maketitle

%%%%%%%%%%%%%%%%%%%%%%%%%%%%%%%%%%%%%%%%%%%%%%%%%%%%%%%%%%%%%%%%%%%%%
%%%%%%%%%%%%%%%%%%%%%%%%%%%%%%%%%%%%%%%%%%%%%%%%%%%%%%%%%%%%%%%%%%%%%
%%%%%%%%%%%%%%%%%%%%%%%%%%%%%%%%%%%%%%%%%%%%%%%%%%%%%%%%%%%%%%%%%%%%%
\section{Introduction}

One  of the fundamental technical problems of the modern microwave/terahertz technology is the development of compact and reliable generators and receivers of coherent electromagnetic signals in the 0.1--10 THz frequency range \cite{Sirtori2002Nat, Kleiner2007Sci, Gulyaev2014JETP}. The THz frequency range has a great potential for applications in medical imaging, security, material characterization, communications, control of technological processes, etc. There are several approaches to THz-frequency generation, including the use of free electron lasers \cite{Nanni2015NatComms}, quantum cascade lasers \cite{Hubers2010Nat}, superconductor Josephson junctions \cite{Ozyuzer2007Sci}, backward wave oscillators \cite{Gorshunov2005BWO}, electro-optic rectification of laser radiation \cite{Tonouchi2007Laser}, etc. However, all the above mentioned sources of THz-frequency signals require rather complex setups, low temperatures, or/and can not be made sufficiently small, which greatly limits their usability in many important practical applications.

Thus, there is a temptation to use novel spin-dependent technologies to generate high-frequency electromagnetic signals. Indeed, the spintronic technology based on the dynamics of spin-polarized electric currents in thin multilayered ferromagnetic structures resulted in the development of spin-torque nano-oscillators (STNO), which are manufactured using e-beam lithography and could be tuned by the variation of both the bias magnetic field or the bias direct current \cite{Chen2016IEEE, Tsoi2000Nat, Kiselev2003Nat, Rippard2004PRL, Demidov2012NatMater, Mohseni2013Science}.
Unfortunately, the frequencies of the signals generated by STNOs are not very high, and, typically, are limited to the interval of 1--50~GHz by the maximum bias magnetic field that can realistically be achieved in a portable spintronic device that uses ferromagnetic materials \cite{Bonetti2009APL}.

It was suggested some time ago \cite{Gomonay2010PRB, Gomonay2014LTP, Cheng2016PRL} that one of the possible ways to substantially increase the frequency of the signals generated in magnetic layered structures is to use in them layers of AFM , that possess a very strong internal magnetic field of the exchange origin which keeps the magnetization vectors of the AFM sublattices antiparallel to each other. Although this idea has been known for quite a while, the realistic theoretical proposal for the development of THz-frequency AFM-based spintronic nano-oscillators has been published only recently \cite{Cheng2016PRL, Khymyn2016SciRep}, soon after the first experimental observation of the switching of AFM sublattices under the action of a DC spin current \cite{Wadley2016Sci, Kriegner2016NatComms}.

In this work we propose and theoretically analyze a THz-frequency signal generator  based on a concept of a ferromagnetic spin-Hall oscillator (SHO) \cite{Demidov2012NatMater, Jungwirth2012NatMater, Demidov2014APL, Chen2016IEEE}, but where SHO free layer is made of a canted  AFM (e.g. Hematite $\alpha$-${\rm Fe}_{2}{\rm O}_{3}$)  that has a small net magnetization $\textbf{m}_{\rm DMI}$ caused by the Dzyaloshinskii-Moriya interaction (DMI). We calculated the electromagnetic power emitted from the antiferromagnetic SHO due to the dipolar radiation from the rotating magnetization $\textbf{m}_{\rm DMI}$ into the free space and into several types of transmission lines and resonators. Our analysis demonstrates that the output power $P_{\rm AC}$ of the AFM-based SHO increases with the generation frequency $f$, and could exceed $P_{\rm AC} = 1 \, \mu$W at $f \sim 0.5$~THz  if a high-quality dielectric resonator is used to receive the generated signal.

\section{Magnetization dynamics induced in AFM by an external spin current}

It was shown previously \cite{Gomonay2010PRB, Gomonay2014LTP, Cheng2016PRL}, that when a layer of an AFM is subjected to an external spin current, e.g. coming from an adjacent current-driven layer of a normal metal (NM) with strong spin-orbital interaction  and polarized along a unit vector $\mathbf{p}$, a corresponding spin-transfer torque (STT) \cite{Berger1996PRB, Slonczewski1996JMMM} is exerted on the sublattice magnetizations of the AFM. This STT  can slightly tilt the AFM sublattice magnetizations $\textbf{M}_{1}$ and $\textbf{M}_{2}$, thus exposing them to the action of a large internal AFM magnetic field of the exchange origin,  which starts to rotate the sublattice magnetizations about the vector $\mathbf{p}$ with a high angular frequency. For the experimentally achievable magnitudes of the spin current the frequencies of this current-induced rotation  lie the THz range.
There is, however, a fundamental problem of how to extract the AC signal corresponding of that THz-frequency rotation from the spin-current-driven AFM, as this extraction is necessary to create a functioning source of a THz-frequency --- AFM-based SHO.

This AC signal can be, in principle, picked up via the spin pumping produced by the rotating AFM magnetizations and inverse spin-Hall effect (ISHE) in the adjacent NM layer. If, however, the chosen AFM is \emph{uniaxial}, the current-induced rotation of the AFM  sublattices is uniform, and the ISHE voltage, which is proportional in this case to the instantaneous angular velocity of the sublattice rotation, is constant and does not contain any high-frequency components.

%If a chosen AFM is \emph{isotropic}, the current-induced rotation of the AFM  sublattices is uniform, and the voltage created by the AFM sublattice rotation in the adjacent layer of a NM due to the inverse spin-Hall effect (ISHE) is proportional to the angular velocity of the sublattice rotation. Thus, this voltage is constant and does contain any high-frequency components.

Two different approaches were used to solve this problem: one could use a feedback-induced nonlinear damping in an AFM \cite{Cheng2016PRL}, or one could use bi-anisotropic AFM, like NiO, that has a relatively strong easy-plane anisotropy and a relatively weak easy-axis anisotropy in the perpendicular direction \cite{Khymyn2016SciRep}. However, both these approaches have serious drawbacks. The first one strongly relies on the quality of the NM/AFM interface, and, also, requires high values of the spin-Hall angle in the NM. The second approach relies on a weak perpendicular anisotropy in the AFM to make the current-induced rotation  of the AFM sublattice magnetizations non-uniform, and, thus, produce a high-frequency component in the voltage signal received in the NM layer through the ISHE \cite{Khymyn2016SciRep}.  Unfortunately, it turned out, that the AC signal created by the ISHE in a NM strongly decreases with the increase of the generation frequency, and  reaches a substantial power ($>1 \mu \text{W}$) only in the range of relatively low frequencies, near the frequency of the in-plane antiferromagnetic resonance ($\sim 100-300$ GHz for NiO SHO \cite{Khymyn2016SciRep}).

\section{AFM canted by the Dzyaloshinskii-Moriya interaction~(DMI)}

\begin{figure}
\centering
\includegraphics[width=0.49\textwidth]{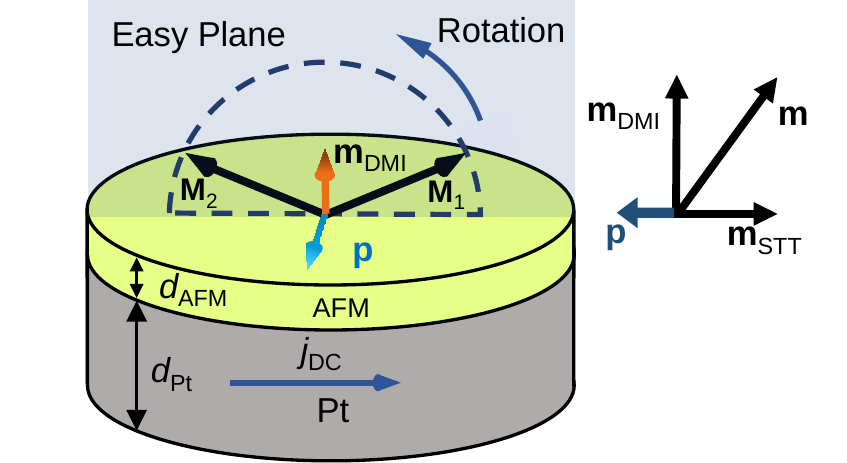}
\caption{
Schematics of an auto-oscillator based on a layered structure containing a Pt  layer of the thickness $d_{\rm Pt}$ and an  AFM layer of the thickness $d_{\rm AFM}$ with sublattice magnetizations  $\textbf{M}_1$ and $ \textbf{M}_2$ canted by the DMI, thus producing a small net magnetization $\textbf{m}_{\rm DMI}$. The rotation of this net magnetization $\textbf{m}_{\rm DMI}$ in the internal exchange field of the AFM layer starts when the perpendicular spin current coming from the Pt layer driven by  a DC electric current of the density $j_{\rm DC}$ tilts the $\mathbf{M}_1$ and $\mathbf{M}_2$ out of the ``easy plane" of the AFM layer.
}
\label{fig:AFM}
\end{figure}

In our current work, we propose a qualitatively different approach to extract the generated AC signal from the AFM layer. We propose to use in the layered structure of the SHO \cite{Khymyn2016SciRep} a \emph{canted} AFM (e.g. Hematite), where magnetic sublattices are canted inside the easy plane by the DMI. This DMI-induced canting results in the formation on a small intrinsic net magnetization vector $\textbf{m}_{\rm DMI}$  of the AFM. It can be shown, that even the uniform rotation of the net magnetization vector $\textbf{m}_{\rm DMI}$ with angular frequency $\omega=2\pi f$  ($f=0.05-2$ THz) leads to a substantial dipolar radiation of a high-frequency signal, that can be received by different types of \emph{external resonators}.

%Therefore, we consider below a bi-layered AFM SHO structure (see Fig. \ref{fig:AFM}) consisting of a layer of a heavy NM  with a strong spin-Hall effect (such as Pt), and a layer of a \emph{canted} AFM insulator with DMI  (such as $\alpha-$Fe$_2$O$_3$) placed in an external resonator.
Therefore, we consider below a bi-layered AFM-based SHO structure (see Fig. \ref{fig:AFM}) consisting of a Pt layer and a Hematite layer placed in an external resonator. For a quantitative estimation of parameters of the proposed AFM-based SHO we used the following typical parameters of the layered structure: it was a circular disk of the radius $r~=~10~\mu$m and thickness $d_{\rm AFM}$ of $5$ nm made of Hematite and covered by a 20~nm thick Platinum layer (see Fig.~\ref{fig:AFM}).

The bulk DMI inside the AFM layer leads to the canting of the magnetizations $\textbf{M}_{1}$ and $\textbf{M}_{2}$ of the AFM sublattices, thus creating  a small net magnetization $\textbf{m}_{\rm DMI} = \textbf{M}_1 + \textbf{M}_2$. The spin dynamics of an AFM with bi-axial anisotropy under the influence of a STT created by an external spin current is described by two coupled Landau-Lifshitz equations for the vectors $\textbf{M}_{1}$ and $\textbf{M}_{2}$:
\begin{multline}
\frac{d\textbf{M}_{ i}}{dt}=\gamma\mu_0\left[ \textbf{H}_{ i}\times\textbf{M}_{i} \right] + \\ + \frac{\alpha_{\rm eff}}{M_{\rm s}} \left[\textbf{M}_{i}\times\ \frac{d\textbf{M}_{ i}}{dt} \right]+\frac{\tau}{M_{\rm s}} \left[\textbf{M}_{i}\times \left[ \textbf{M}_{i}\times\textbf{p}\right] \right],
\label{eq:LL}
\end{multline}
where $i=1,2$ are the indices denoting the AFM sublattices, $\gamma$ is the modulus of the gyromagnetic ratio, $\mu_0$ is the vacuum permeability, $M_{\rm s}$ is the magnitude of saturation magnetization of the AFM sublattices, $\alpha_{\rm eff}$ is the effective AFM Gilbert damping parameter, $\tau$ is the amplitude of the STT caused by the spin current transferred from the current-driven Pt layer into the AFM layer, $\mathbf{p}$ is a unit vector along the spin current polarization, and $\mathbf{H}_{i}$ is the effective magnetic field acting on the sublattice $\mathbf{M}_{i}$:
\begin{multline}
\textbf{H}_{1,2} = (- H_{\rm ex} \textbf{M}_{2,1}
- H_{\rm h} \textbf{n}_{\rm h} \left( \textbf{n}_{\rm h} \cdot \textbf{M}_{1,2}  \right) +
\\
+ H_{\rm e} \textbf{n}_{\rm e} \left( \textbf{n}_{\rm e} \cdot\textbf{M}_{1,2}\right) \mp H_{\text{DMI}}\left[ \textbf{n}_{\rm DMI} \times\textbf{M}_{2,1} \right])/M_{\rm s}.
\label{eq:fields}
\end{multline}

Here $H_{\rm ex}$ is the exchange field, $H_{\rm e}$ and $H_{\rm h}$ are the in-plane and perpendicular-to-plane anisotropy fields, respectively, $\textbf{n}_{\rm e}$ and $\textbf{n}_{\rm h}$  are the unit vectors along the ``easy" and ``hard" anisotropy axes, $H_{\rm DMI}$ is the effective field caused by the DMI, and $\textbf{n}_{\rm DMI}$ is the DMI vector.

The STT amplitude expressed in the frequency units \cite{Nakayama, Khymyn2016SciRep} is:
\begin{equation}
%%\label{eq:tau}
\tau
=j_{\rm DC} g_{\uparrow \downarrow} \theta_{\text{SH}} \frac{e \gamma\lambda\rho_{\rm Pt} }{2\pi M_{\rm s} d_{\text{AFM}}} \tanh{\frac{d_{\text{Pt}}}{2\lambda_{\rm Pt}}}
\end{equation}	
where $j_{\rm DC}$ is the density of the DC electric current in the Platinum layer,  $g_{\uparrow \downarrow}$ is the spin-mixing conductance at the Pt/AFM interface, $\theta_{\rm SH}$ is spin-Hall angle in Pt, $e$ is the modulus of the electron charge, $\lambda$ is the spin-diffusion length in the Pt layer, $\rho_{\rm Pt}$  is the electric resistivity of Pt,   $d_{\rm AFM}$ and $d_{\rm Pt}$ are the thicknesses of the AFM and Pt layers, respectively.

The effective damping parameter $\alpha_{\rm eff}$ of the layered structure in Eq.(\ref{eq:LL}) includes the additional magnetic losses due to the spin pumping from the AFM layer into the adjacent Pt layer:
\begin{equation}
\label{eq:alfa}
\alpha_{\rm eff}
=
\alpha_0 +  g_{\uparrow \downarrow} \frac{\gamma\hbar}{4\pi M_{\rm s} d_{\rm AFM}},
\end{equation}		
where $\alpha_0$ is intrinsic Gilbert damping constant,  $\hbar $ is the reduced Planck constant.

As it was shown in \cite{Khymyn2016SciRep}, the presence of the anisotropy $H_{\rm e}$ in the plane perpendicular to the spin-polarization direction $\mathbf{p}$ makes the current-driven magnetization dynamics in AFM nonuniform in time, and, also determines the threshold charge current at which the auto-oscillatory dynamics starts. Thus, to minimize the threshold of auto-oscillations one should choose almost purely ``easy-plane" AFM with $\mathbf{n}_{\rm h} \parallel \mathbf{p}$ and low value of the perpendicular anisotropy ($H_{\rm e} \ll H_{\rm h}$). The DMI vector $\mathbf{n}_{\rm DMI}$ is, commonly, directed along one of the AFM crystallographic axes. If $\mathbf{n}_{\rm DMI} \parallel \mathbf{n}_{\rm h}$, the DMI creates a small net magnetization, which lies in the easy plane of the AFM, $\mathbf{m}_{\rm DMI} \perp \mathbf{n}_{\rm h}$. The above described geometrical relations are realized, for example, in  $\alpha$-${\rm Fe}_{2}{\rm O}_{3}$ (Hematite) and in $\text{FeBO}_3$.

Below, we consider in detail the SHO based on a thin film of Hematite, which has almost purely easy-plane anisotropy ($H_{\rm e}=0.2 \text{ Oe}= 15.9~\rm{A/m} $ while $H_{\rm h}=200~\text{ Oe}= 15.9\cdot 10^3~ \rm{A/m}$, $H_{\rm ex}=9 \cdot 10^6 \text{ Oe}=0.7\cdot 10^9~\rm{A/m}$) and net magnetization moment $m_{\rm DMI}=M_{\rm s} H_{\rm DMI}/H_{\rm ex}= 2100 \text{ A/m}$ caused by the $H_{\rm DMI}= 22\cdot 10^3 \text{ Oe}=1.75\cdot 10^6~\text{ A/m}$ \cite{Anderson, hematiteBook, TurovBook, Kumagai}. The intrinsic magnetic damping of Hematite is rather low $\alpha_0 \simeq 10^{-4}$, while the effective damping that takes into account the spin pumping into the adjacent Pt layer (see Eq. (\ref{eq:alfa})) is $\alpha_{\rm eff}=2 \cdot10^{-3}$.

We solved Eq. (\ref{eq:LL}) numerically for the case when the spin current polarization $\mathbf{p} \parallel  \mathbf{n}_{\rm h}$ with the main material and geometric parameters taken from \cite{Khymyn2016SciRep}. In this case, the STT (i.e. the last term in Eq. (\ref{eq:LL})) tilts the AFM sublattice magnetizations $\textbf{M}_{1}$ and $\textbf{M}_{2}$ out of the easy plane of the AFM, and creates the net magnetic moment
\begin{equation}
m_{\rm STT}=\frac{\tau M_{\rm s}}{\alpha_{\rm{eff}} \gamma H_{\rm ex}}
\end{equation}
in the out-of-plane direction $\mathbf{m}_{\rm STT} \parallel \mathbf{p} \parallel \mathbf{n}_{\rm h}$. Now, the total net magnetization reads as $\mathbf{m}=\mathbf{m}_{\rm DMI}+\mathbf{m}_{\rm STT}$ and $\mathbf{m}_{\rm DMI}\perp \mathbf{m}_{\rm STT}$.

The $\textbf{M}_{1}$ and $\textbf{M}_{2}$ and, consequently, $\mathbf{m}_{\rm DMI}$, which are now exposed to the action of the internal exchange field, start to rotate around $\mathbf{p}$ with the angular velocity $2 \pi f=\tau/\alpha_{\rm eff}$. This rotation is almost uniform in time, due to the low value of the in-plane anisotropy in Hematite.

\begin{figure}
\centering
\includegraphics[width=0.49\textwidth]{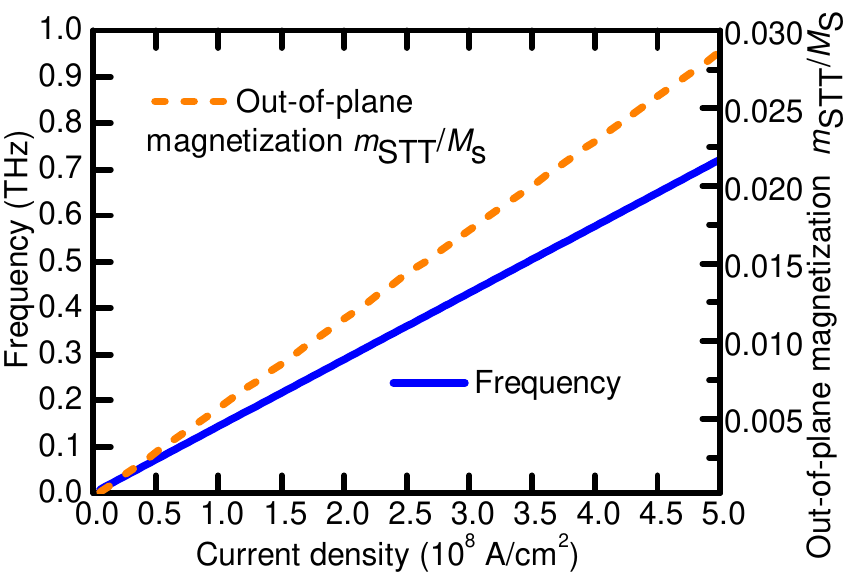}
\caption{
Calculated frequency of the generated AC signal $f$ (solid line, left axis) and  normalized magnitude of the out-of-plane magnetization $ m_{\rm STT}/ M_{\rm s}$ (dashed line, right axis) in an SHO based on a canted AFM (Hematite) as functions of the DC driving current density $j_{\rm DC}$ in the adjacent Pt layer for the Hematite film thickness of $d_{\rm AFM} = 5$~nm.
}
\label{fig:fmJ}
\end{figure}

The rotation frequency  $f$, as well as the net magnetization $m_{\rm STT}$, are shown as functions of the electric current density flowing in the Pt layer in Fig. \ref{fig:fmJ}. To obtain these curves it is necessary to know the value of the spin-mixing conductance $g_{\uparrow \downarrow}$ at the Pt/Hematite interface.  Although there exist several studies of the electronic structure of Hematite \cite{hemElec1, hemElec2, hemElec3}, we did not find a reliable value of  $g_{\uparrow \downarrow}$ in literature. For numerical simulations we choose a value $g_{\uparrow \downarrow} = 6.9 \cdot 10^{14} \text{cm}^{-2}$, which was previously obtained for the Pt/NiO interface \cite{ChengSpinMix} and is in a good agreement with the rough estimations made by the method developed in \cite{Xiao2010}. The assumption for the $g_{\uparrow \downarrow}$ magnitude also allows us to make a direct comparison of both the general properties  and the intrinsic spin dynamics for the SHO based on Hematite and a similar auto-oscillator based on the NiO \cite{Khymyn2016SciRep}.

As one can see from Fig.~\ref{fig:fmJ}, the generation frequency $f$ can be controlled by the density $j_{\rm DC}$ of the electric current injected in the Pt layer. For instance, the current density $j_{\rm DC}$ that is required to get the generation at the frequency of $f = 0.5$~THz is $j_{\rm DC}=3.5\cdot10^{8}$ A/cm$^2$, which has been previously achieved in experiment \cite{Demidov2012NatMater}.

\section{Dipolar electromagnetic radiation of a rotating net magnetization in a canted AFM}
To investigate the possibility of power extraction from an SHO based on an Pt/AFM layered structure, let us, first, consider the electric current in a Pt layer produced by the ISHE caused by a current-driven rotation of the AFM sublattice magnetizations. The density of the spin current flowing back from the AFM layer to the adjacent Pt layer $\mathbf{j}_{\rm s}^{\rm out}$ can be written as:
\begin{equation}
 \mathbf{j}_{\rm s}^{\rm out}=\frac{\hbar g_{\uparrow \downarrow}}{4 \pi M_{\rm s}^2}  \sum_{i=1,2} \left(\mathbf{M}_{ i} \times  \mathbf{\dot M}_{ i} \right).
 \label{eq:outcurr}
\end{equation}

In our case of an easy-plane AFM with DMI, the total net magnetization caused by the DMI is precessing along  a cone  of the height $m_{\rm STT}$ and cone base radius $m_{\rm DMI}$ (see Fig. \ref{fig:AFM}). Therefore, the AC component of the current $\mathbf{j}_{\rm s}^{\rm out}$ (see~Eq.~(\ref{eq:outcurr})) is proportional to the product of \emph{two small quantities }$\sim  m_{\rm DMI} m_{\rm STT}$ and, consequently, in the considered frequency range it is negligibly small (see Fig. \ref{fig:fmJ}). In particular, we obtain the maximum output power less than $ 100~\text{pW}$ at $f = 1 \text{ THz}$ for a square Hematite sample having $200~\rm{\mu m}^2$ surface area. Such a low value of the output power makes the ISHE practically useless as a method of the signal extraction in the frequency range $2\pi f~ \ll~ \gamma\mu_0 H_{\rm ex}$, and it is necessary to find other ways to extract an output AC signal from an SHO  based on a canted AFM.

Fortunately, in a canted AFM crystal, having a small net magnetization, the current-induced precession of this magnetization can be detected not only through the ISHE, but also \emph{directly} through the dipolar radiation produced by this precessing magnetization. The rotating magnetization of the AFM-based SHO $\textbf{m}_{\rm DMI}$ creates an oscillating dipolar magnetic field that can be received, channeled, and, then, utilized, if the generating SHO is coupled to an appropriate resonator.

The problem of the direct dipolar emission of an AC signal generated by a precessing magnetization has been considered in \cite{Prokopenko2011ML} for the case of a conventional microwave-frequency STNO. It was shown in \cite{Prokopenko2011ML} that the dipolar emission mechanism might become preferable for the case of magnetic devices operating at frequencies above 0.1~THz, which makes this mechanism promising for application in the THz-frequency AFM-based SHOs.

To calculate the AC power $P_{\rm AC}$ that can be emitted by the SHO into a free space, different transmission lines (rectangular waveguide, parallel plate waveguide, dielectric waveguide) and different resonators (rectangular, parallel plate, dielectric) we used a simple model of a direct dipolar emission from a system of two effective magnetic dipoles developed in \cite{Amin2009TMagn, Prokopenko2011ML}. In the framework of this approach we used the expressions for the fields of a magnetic dipole obtained in \cite{Amin2009TMagn, Ramo1984} and the standard expressions for the electromagnetic fields of fundamental modes in the considered transmission lines and resonators presented in \cite{Ramo1984}. Also, to simplify the theoretical analysis of the electromagnetic field excitation in a rectangular dielectric waveguide and resonator by a net AC magnetization $\textbf{m}_{\rm DMI}$ we used the approximate magnetic wall boundary conditions \cite{Ramo1984}. Unfortunately, the exact analytical solution for this problem has not been found, so far.

In our approximate calculation we assumed that the rotating net magnetization of the AFM-based SHO  $\textbf{m}_{\rm DMI}$ is  spatially uniform (macrospin approximation),  and that the sizes of the effective magnetic dipoles (defined by the in-plane dimensions  of the SHO) are substantially smaller than the wavelength of the AC signal. To evaluate the maximum magnitude of the emitted AC power we took the magnitude of the AC magnetization $\textbf{m}_{\rm DMI}$ from the numerical solution of Eq. (\ref{eq:LL}).

\begin{table*}
\caption{Expressions for $V_{\rm eff}$ and values for the
AC power emitted by an SHO at $f=0.5$~GHz calculated using
Eq.~(\ref{eq:Pac})} \label{Tbl}
\begin{tabular}{lccc}
	\hline
    \hline	
    Case & Expression for $V_{\rm eff}$ & Parameters & Maximum power, W\\
	\hline
	Free space &
		$3 c^3/8\pi^3 f^3$ &
		$f = 0.5 \ {\rm THz}$, $c = 3\cdot 10^8 {\rm m/s} $, $Q = 1$ &
		$2.6 \cdot 10^{-12}$ \\
	Nano-loop &
		$r^2 R_{\rm L}/2\mu_{0}\pi^2 f$ &
		$R_{\rm L} = 6030 \ \Omega$,  $Q = 1$ &
		$1.4 \cdot 10^{-10}$ \\
	Rectangular waveguide &
		$2 a^2 b \chi / \pi^2$ &
		$a = 0.47 \ {\rm mm}$, $b = 50 \ {\rm nm}$,
		$\chi = \eta / \sqrt{1 - \eta^2}$, &
		$2.3 \cdot 10^{-8}$ \\
		&
		&
		$\eta = c / 2 a f \approx 0.64$, $Q = 1$ & \\
    Dielectric waveguide &
		$2 a^2 b \chi_{\varepsilon} / \pi^2$ &
		$a = 0.47 \ {\rm mm}$, $b = 50 \ {\rm nm}$,
		$\chi_{ \varepsilon} = \eta_{\varepsilon} / \sqrt{1 - \eta_{\varepsilon}^2}$, &
		$2.9 \cdot 10^{-8}$ \\
		&
		&
		$\eta_{\varepsilon} = c / 2a f\sqrt{\varepsilon} \approx 0.2$, $\varepsilon =10$, $Q = 1$& \\
	Parallel plate line &
		$8 a b c / \pi^2 f$ &
		$a = 0.47 \ {\rm mm}$, $b = 50 \ {\rm nm}$, $Q = 1$ &
		$1.2 \cdot 10^{-9}$\\
	Rectangular resonator &
		$ 8 a^4 b \chi^3 f^2/ c^2  $ &
		$a = 0.47 \ {\rm mm}$, $b = 50 \ {\rm nm}$, $Q = 10$ &
		$2.26 \cdot 10^{-9}$\\
	Parallel plate resonator &
		$ 2 a b c/f $ &
		$a = 0.47 \ {\rm mm}$, $b = 50 \ {\rm nm}$, $Q = 2$ &
		$9.7 \cdot 10^{-10}$\\
    Dielectric resonator &
		$ 2 a^2 b \chi_{\varepsilon} $ &
		$a = 0.47 \ {\rm mm} $, $b = 50 \ {\rm nm}$, $\varepsilon =10$,  $Q = 750$ &
		$1.1 \cdot 10^{-6}$\\
	\hline
    \hline
\end{tabular}
\end{table*}

Using the above described model we obtained the following generalized expression for the maximum AC power that is emitted by an AFM-based SHO in \emph{any} of the microwave/THz-frequency resonators (or wave\-guides) coupled to the SHO:
\begin{equation}
\label{eq:Pac}
    P_{\rm AC} = P_{\rm m} \frac{V}{V_{\rm eff}} Q \ .
\end{equation}
Here $P_{\rm m} = \mu_0 m_{\rm DMI}^2 V f$ is the characteristic AC power generated in the SHO by the rotating magnetization $\textbf{m}_{\rm DMI}$, $f$ is the frequency of the generated AC signal, $V=\pi r^2 d_{\rm AFM}$ is the volume of the AFM layer,
$V_{\rm eff}$ is the effective frequency-dependent volume of a particular resonator coupled to SHO and $Q$ is the frequency-dependent quality factor of this resonator.

It follows from Eq.~(\ref{eq:Pac}) that a significant output AC power $P_{\rm AC}$ can be obtained only when the resonator coupled to the AFM-based SHO has a reasonably high quality factor ($Q \gg 1$), and the SHO is operating at a sufficiently high frequency, as $P_{\rm m}$ \emph{increases} with the  increase of the generation frequency $f$.  This generation frequency can be controlled by the magnitude of the driving current density (see Fig.~\ref{fig:fmJ}). It is, also, important to have a sufficiently small ``effective volume" $V_{\rm eff}$ of the SHO resontor, which, in the ideal case, should be comparable to the volume $V$ of the  AFM layer.

The extracted AC power should be substantially smaller in the case of radiation into a transmission line, where $Q = 1$, and/or the effective volume $V_{\rm eff}$ is, typically,  substantially larger than the volume of the AFM layer.

To decrease the effective volume of the resonance system one could fill it with a dielectric having a large dielectric permittivity $\varepsilon$.  This approach is well-known in the microwave and terahertz-frequency technology \cite{Ramo1984}.
A simple qualitative analysis shows that a high-Q THz-frequency dielectric resonator could be an effective system for the extraction of the generated AC power from an AFM-based SHO. It is also obvious, that the intrinsic AC power $P_{\rm m} \sim fV$, and the efficiency of the AC power extraction, using the dipolar radiation mechanism, are higher for the AFM-based SHOs, than for the conventional ferromagnetic STNO. The reasons for that are, first of all, the much higher frequencies $f$ generated by AFM-based SHOs  ($f \sim 1$~THz in an SHO, while it is $f \sim 10$~GHz in a typical STNO, see Fig.~\ref{fig:fmJ}), and a substantially larger active magnetic layer volume $V$ in an AFM-based SHO, compared to an STNO (typical radius of a conventional circular STNO is about 100~nm, while the radius of  an antiferromagnetic SHO could be 100 and more times larger).

Despite a relatively small magnitude of the net magnetization $\textbf{m}_{\rm DMI}$  rotating inside a canted AFM and, consequently, a relatively small magnitude of  the intrinsic power $P_{\rm m}$, one could obtain a substantial output power $P_{\rm AC}$ of the total auto-generator based on layer of a canted AFM in the case of a sufficiently  high generation frequency $f$.

\begin{figure}
\centering
\includegraphics[width=0.49\textwidth]{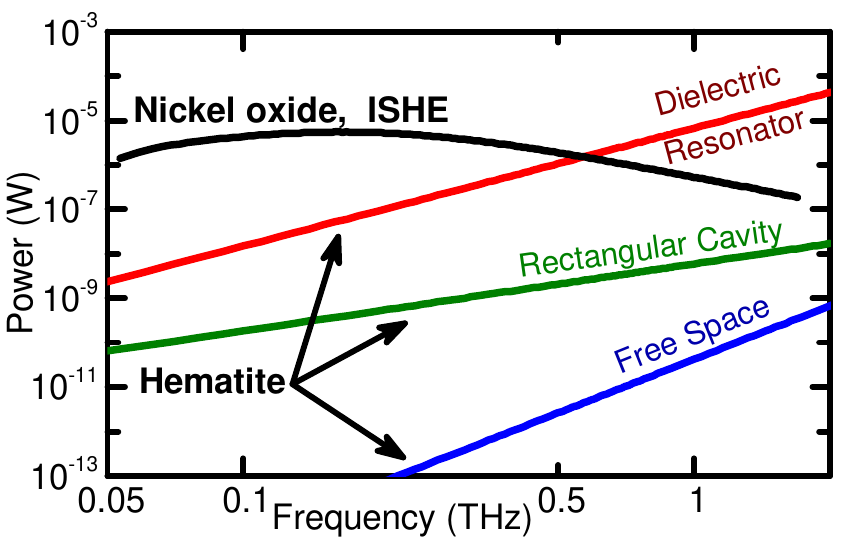}
\caption{
Generated power vs. frequency for an SHOs based on a layer (thickness $d_{\rm AFM}=5$~nm) of a canted AFM providing dipolar radiation into different types of resonance systems: dielectric resonator (red line), rectangular cavity (green line), free space (blue line). For comparison, a similar curve is presented for an SHO based on a layer of bi-anisotropic AFM (NiO) where the generated AC signal is extracted through the ISHE in the adjacent Pt layer (black  line demonstrating decrease of the generated power with frequency).
}
\label{fig:Pf}
\end{figure}

\begin{figure}
\centering
\includegraphics[width=0.49\textwidth]{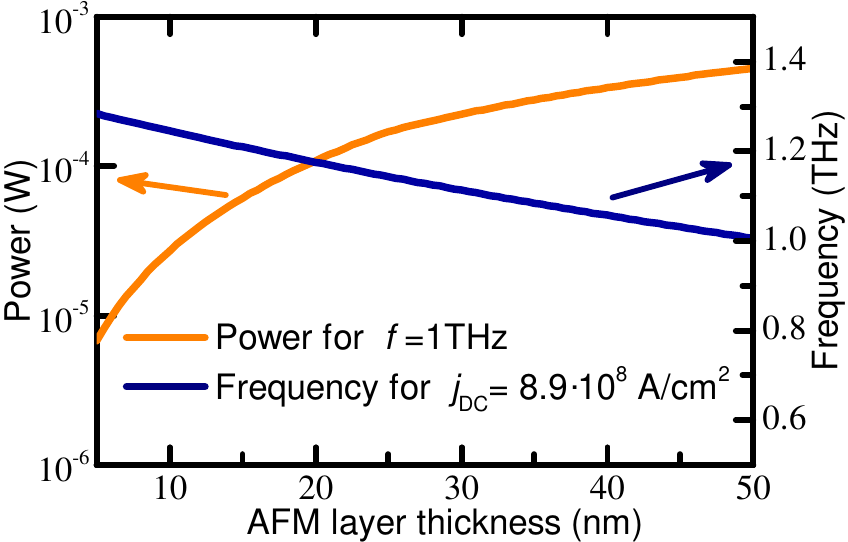}
\caption{
Power and frequency  of a signal generated in an SHO based on a canted AFM (Hematite) as functions of the AFM layer thickness. The power was calculated for the generated frequency  $f$ = 1 THz, while the frequency was calculated  at the driving DC current
$j_{\rm DC} = 8.9 \cdot 10^8 \rm{A/cm^2}$.
}
\label{fig:THICK}
\end{figure}

Using Eq.~(\ref{eq:Pac}) we calculated the maximum AC power $P_{\rm AC}$ radiated  by a Hematite-based SHO into different resonance structures (free space, rectangular and dielectric resonator) as functions of the generated frequency. These curves are shown in Fig.~\ref{fig:Pf}. In Fig.~\ref{fig:Pf} blue line shows the dependence $P_{\rm AC}(f)$ for a Hematite SHO ($d_{\rm AFM} =5$~nm) radiating into  free space, while the green line and red line show the dependence $P_{\rm AC}(f)$ for the Hematite-based SHO coupled to a rectangular cavity and a high-Q dielectric resonator, respectively.
For comparison, a similar curve (black line) is presented for an SHO based on a layer of a bi-anisotropic AFM (NiO) ($d_{\rm AFM} =5$~nm),  where the generated AC signal is extracted through the ISHE in the adjacent Pt layer. It is clear, that the signal extraction method based on the ISHE has an advantage at relatively low frequencies, but the method based on the dipolar radiation  from  a canted AFM wins in the limit of high  THz  frequencies.

The dependences of the power and frequency  of a signal generated in an SHO based on a canted AFM (Hematite) on the AFM layer thickness are presented in Fig.\ref{fig:THICK}. The power was calculated for the generated frequency  $f$~=~1~THz, while the frequency was calculated  at the driving DC current $j_{\rm DC} = 8.9 \cdot 10^8 \ \rm{ A/cm^2} $.  It is clear that the increase of the AFM layer thickness leads to the moderate decrease of the generated frequency, but to a substantial increase of the power  of the generated THz-frequency signal.

The general expressions for the effective volume $V_{\rm eff\,}$ along with the system's parameters and the values of $P_{\rm AC}$ (calculated at a generation frequency of $f=0.5$~THz) are presented in Table~\ref{Tbl}.

The results presented in the Table~\ref{Tbl} and Fig.~\ref{fig:Pf} (blue line) demonstrate that the power emitted from an AFM-based SHO operating at a frequency $f = 0.5$~THz into a free space is very low. If we place an SHO in the center of a gold wire nano-loop of a round shape (radius $r_{\rm L} = 100~\mu$m, square cross-section of the wire is $S_{\rm L} = 50\times50 \ n{\rm m}^2$, and the characteristic resistance $R_{\rm L}~=~2\pi \rho_{\rm Au}  r_{\rm L}/S_{\rm L} = 6030\ \Omega,$ where $\rho_{\rm Au} = 24 \ n \Omega \cdot{\rm m}$ is the resistivity of gold) we can increase the emitted power by approximately $10^2$ times, because the power can be collected by the loop in a near-field zone (see Table ~\ref{Tbl})\cite{Prokopenko2011ML}.

To increase the generated power we can also place an AFM-based SHO in a waveguide (rectangular, parallel plate or dielectric).
In this case the electromagnetic field generated by an SHO can excite  fundamental propagating modes in these transmission lines, but, as it can be seen from Table~\ref{Tbl}, this approach is not very effective, because for the considered waveguides $Q= 1$, and $V_{\rm eff} \gg V$, which, in accordance to Eq.~(\ref{eq:Pac}) leads to rather small values of $P_{\rm AC}$.

In order to substantially enhance the emitted power, an SHO should be placed in a microwave/THz-frequency resonator with a sufficiently high $Q$, which allows one to increase the emitted power $Q$ times (see Eq.~(\ref{eq:Pac})).
Our calculations performed for rectangular, parallel plate and dielectric resonators having sizes $a \times b \times l$ and reasonably high frequency-dependent quality factors $Q \equiv Q(f)$\cite{Qf} demonstrated, that  power emitted into a rectangular or parallel plate resonators having metal walls is comparable to the power that can be extracted from an SHO placed in transmission line, mainly due to the increase of the ohmic losses in the resonator walls (both resonators) and radiation losses (parallel plate resonator).

The power emitted into a resonator can be increased if the effective volume $V_{\rm eff}$ of the resonator is reduced, while its quality factor remains sufficiently large. To achieve this, it is possible e.g. to place an SHO inside a dielectric resonator having the resonance frequency $f=0.5$~THz (mode ${\rm TM}_{101}$), reasonable sizes ($470 \, \mu{\rm m} \times 50 \, {\rm nm} \times 97 \, \mu{\rm m}$), and the dielectric permittivity $\varepsilon = 10$ and Q-factor $Q=750$.  In such a case the power emitted by the AFM-based SHO into a dielectric resonator could reach $P_{\rm AC} = 1.1 \, \mu$W (see Table~\ref{Tbl}, and a solid line in Fig.~\ref{fig:Pf}). It is clear from  Fig.~\ref{fig:Pf} and Table~\ref{Tbl}, that the design of an AFM-based SHO involving a high-Q dielectric resonator could be promising for the development of practical THz-frequency AC signal sources based on the antiferromagnetic SHOs.

At the same time, at the frequencies higher than 1~THz the use of quasi-optical resonators might turn out to become preferable.
Also, as it is clear from Fig. \ref{fig:THICK}, an additional enhancement of the AC power $P_{\rm AC}$ emitted from an AFM-based SHO could be achieved by increasing the thickness $d_{\rm AFM}$ of the AFM layer.  This will lead to the increase of the AFM layer volume $V \sim d$ and, therefore, to the increase in the power of magnetization oscillations $P_{\rm m}$ (see Eq.~(\ref{eq:Pac})).

Finally, it is interesting, for comparison, to consider a bi-anisotropic AFM crystal with no DMI (e.g. nickel oxide (NiO)) where the AC component of the output spin current can reach a substantial magnitude, if the AFM sublattice rotation is non-uniform in time \cite{Khymyn2016SciRep}. In this case, the AC component of the current $\mathbf{j}_{\rm s}^{\rm out}$ is proportional to the acceleration in the rotation of  the sublattice magnetizations. The SHO based on this effect was proposed in \cite{Khymyn2016SciRep}, where the small in-plane anisotropy of the NiO makes the sublattice rotation non-uniform in time. The calculated output power of such an NiO SHO is shown on Fig. \ref{fig:Pf} by a black solid line. To make a direct comparison with the case of an ``easy-plane" Hematite SHO we assumed that both types of AFM-based SHOs  have the same surface area. As one can see from  Fig. \ref{fig:Pf}, the output power of the NiO SHO decreases with the increase of the generation frequency, and this device and this method of the AC signal extraction become non-competitive for generation frequencies above $0.5$ THz.

\section{Conclusions}
In conclusion, we have demonstrated theoretically that an SHO based on a canted antiferromagnet (e.g. Hematite) could be used for the development of THz-frequency AC signal sources, where the power of a generated AC signal can be extracted from an SHO through the dipolar oscillating magnetic field created by the current-driven rotating net magnetization of the canted AFM.
We have shown that the efficiency of this mechanism of the AC power extraction is increased with the increase of the signal frequency, and depends on the Q-factor and the effective volume of the attached microwave/THz-frequency resonance system.
Our analysis has also shown, that the output AC power of such a THz-frequency signal source could exceed 1~$\mu$W at the frequency $f \sim 0.5$~THz for the Hematite/Pt SHO coupled to a dielectric resonator with reasonable experimental parameters (sizes $470 \, \mu{\rm m} \times 50 \, {\rm nm} \times 97 \, \mu{\rm m}$, dielectric permittivity $\varepsilon = 10$ and Q-factor $Q=750$). The above proposed SHO based on a  current-driven layered structure  of a canted AFM and Pt and incorporating an external dielectric resonator has a practically interesting level of the output power, and its efficiency increases with the increase of the generation frequency, in contrast to the NiO SHO, proposed in \cite{Khymyn2016SciRep}, which relies on the ISHE mechanism for the extraction of the AC power. The obtained results could become critically important for the development and optimization of practical THz-frequency nano- and micro-scale electromagnetic signal sources.

\section{Acknowledgements}

 This work was supported by the Grants Nos. EFMA-1641989 and ECCS-1708982 from the NSF of the USA, the Knut and Alice Wallenberg foundation (KAW), by the Center for NanoFerroic Devices (CNFD) and the Nanoelectronics Research Initiative (NRI), by DARPA, by the Taras Shevchenko National University of Kyiv, Ukraine (Grant No. 16BF052-01), and by the National Academy of Sciences of Ukraine (Grant No. 7F). B.A.I was partly supported by the National Academy of Sciences of Ukraine via the project \#. 1/17-N. The publication also contains results of the studies conducted under the ``President’s of Ukraine'' grant for competitive projects (F74/25621) of the State Fund for Fundamental Research of Ukraine.

%%%%%%%%%%%%%%%%%%%%%%%%%%%%%%%%%%%%%%%%%%%%%%%%%%%%%%%%%%%%%%%%%%%%%


\begin{thebibliography}{99}
\bibitem{Qf}
We choose the qualitative approximation for the frequency dependent Q-factor as $Q(f)=Q_0 \cdot (f_0/f)^{3/2}$ for the rectangular ($Q_0 = 10^5$ at $f_0 = 1$~GHz), $Q(f)=Q_0 \cdot (f_0/f)$ for the parallel plate ($Q_0 = 10^3$ at $f_0 = 1$~GHz) resonators, and $Q(f)= Q_0 \cdot (f_0/f)^{1/3}$ for the dielectric resonator ($Q_0 = 6 \cdot 10^3$ at $f_0 = 1$~GHz).
\end{thebibliography}
\end{document}